\documentclass[twocolumn,pra,amsmath,amssymb,superscriptaddress,nofootinbib]{revtex4-2}

\usepackage[utf8]{inputenc}
\usepackage[T1]{fontenc}
\usepackage{hyperref}
\usepackage{url}
\usepackage{xcolor}
\usepackage{graphicx}
\usepackage{tikz}
\usetikzlibrary{arrows.meta}

\hypersetup{
  colorlinks=true,
  linkcolor=blue!60!black,
  citecolor=blue!60!black,
  urlcolor=blue!60!black
}

\begin{document}

\title{Market Dynamics, Governance and Open Research Metadata in the AI Era}

\author{Daniel~W.~Hook}
\email{d.hook@digital-science.com}
\affiliation{Digital Science, 6 Briset Street, London, EC1M 5NR, UK}

\date{March 2026}

\begin{abstract}
The debate about scholarly knowledge infrastructure has long been framed as a contest between openness and commercial enclosure. This framing distorts both policy and practice. The real tension lies between the persistent cost of producing and refining structured metadata under deep technological friction, and the differentiated demands distinct communities place on data quality, focus and granularity. We introduce the \emph{innovation annulus}: the zone between freely available structured data and the advancing frontier of commercially refined knowledge products. This zone is a permanent, functional feature of the ecosystem---not a pathology to eliminate. By analogy with the efficient market hypothesis, its width measures production inefficiency, set by the interplay of friction and demand. Artificial intelligence reshapes the annulus, lowering barriers to basic structuring, raising the threshold at which refinement adds value, and introducing systemic risks through unprovenanced AI-derived metadata. CRediT contributions, funding acknowledgements and AI disclosure statements illustrate the annulus lifecycle. Governance should calibrate the annulus, not abolish it: thin enough to serve research efficiently, wide enough to sustain innovation. A formal welfare framework, analogous to the Nordhaus optimal patent life, characterises the trade-offs and yields testable predictions. The Barcelona Declaration offers a promising forum for boundary governance.
\end{abstract}


\maketitle

\section{Introduction}
\label{sec:introduction}

For more than three decades, the debate about scholarly knowledge infrastructure has been organised around an opposition between openness and commercial enclosure. On one side, a sustained community of advocates, funders and policymakers has argued that the broadly defined products of publicly funded research including metadata about this research should be openly available~\cite{Suber2012}. On the other, commercial actors have maintained structured data products whose value depends, to varying degrees, on restricting access. The result has been a policy conversation framed as a zero-sum contest: Every advance in openness is a retreat for commercial interests, and vice versa.

This framing has been productive in some respects. It has driven the creation and development of open identifier infrastructure---Crossref, ORCID, ROR, DataCite---and the progressive opening of citation data and abstract data through initiatives such as I4OC and I4OA respectively. It has generated mandates for open access to published research and, most recently, the Barcelona Declaration on Open Research Information~\cite{BarcelonaDeclaration2024}, which extends the logic of openness from publications to the broader research metadata ecosystem. These are significant achievements.

Yet this framing, at a deeper level, is now less productive---and its persistence has distorted both policy and practice \cite{Khoo2019,Butler2023,Debat2020}. We assert that the central tension in scholarly knowledge infrastructure is not between openness and commerce. It is between two features of the system that will not go away: First, the persistent and non-trivial cost of producing, structuring and refining knowledge data in a system characterised by deep technological frictions; and second, the demands that a variety of communities of users have for data to support a radically differentiated use cases that require vastly different data quality, focus and granularity.

\begin{figure*}[t]
\centering
\includegraphics[width=\textwidth]{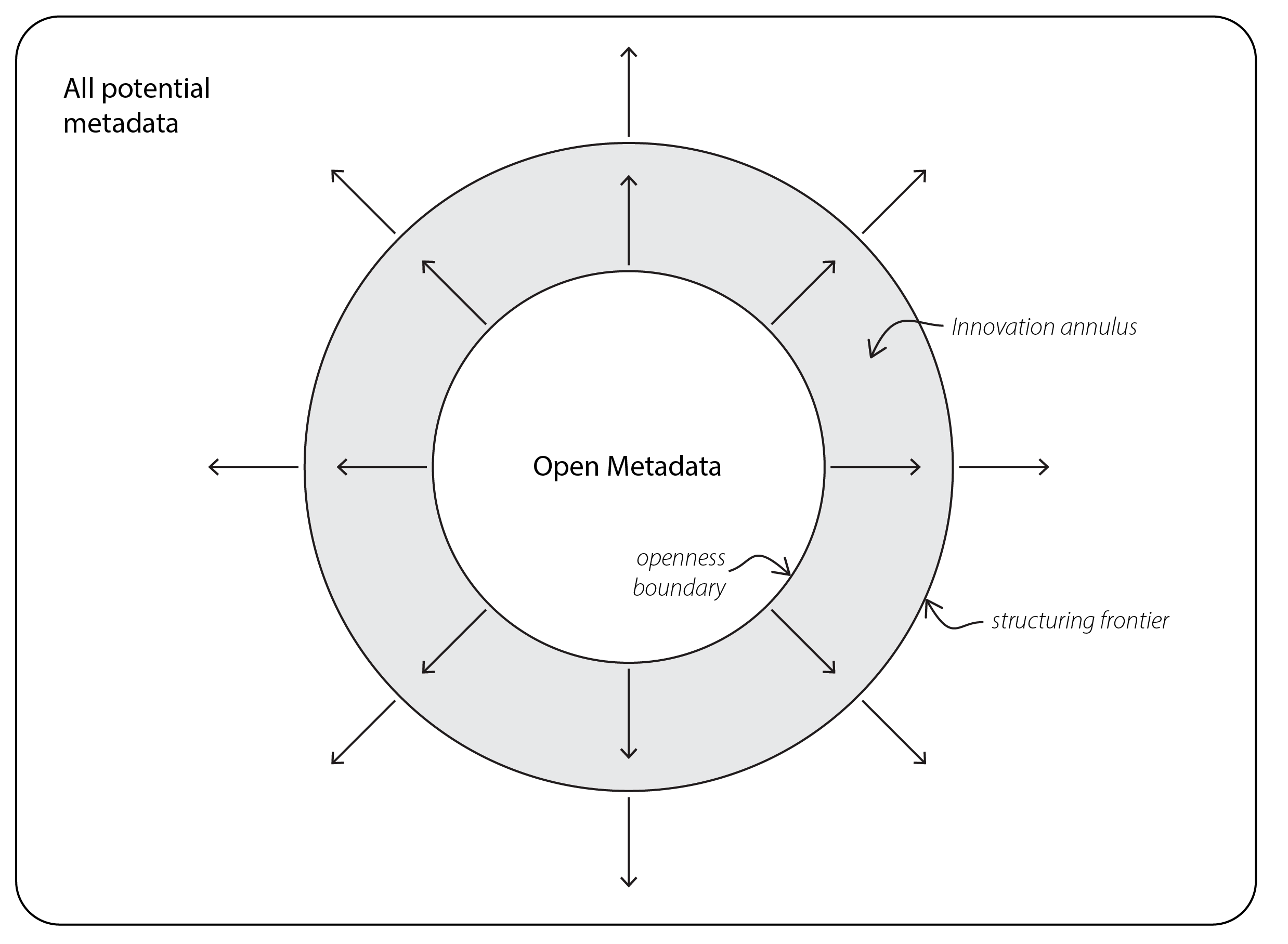}
\caption{\label{fig1}\small\textbf{The Innovation Annulus.} The outer circle represents the \emph{structuring frontier}---the frontier of what has been structured and made available in analysable form. Beyond the outer circle lies unprocessed, unstructured, or as-yet-unpublished scholarly output. The inner circle represents the boundary of \emph{openness boundary}---data that is freely licensed, transparently governed, stable and available for unrestricted reuse. The zone between the two circles is the \emph{innovation annulus}: Data that has been structured and refined to a level that makes it useful, but that has not yet been commoditised to the point of free availability. Arrows on the outer circle indicate expansion (new data types, new structuring capabilities). Arrows on the inner circle indicate outward migration (standards adoption, collective disclosure, technological commoditisation). The width of the annulus at any point represents the gap between the current frontier of commercially refined data and the current baseline of open provision. \textit{Adapted from Hook~\cite{Hook2024blog}}.}
\end{figure*}

These two features interact to produce what we call the \emph{innovation annulus}~\cite{Hook2024blog}---a zone between the open core of structured data that is either free or indistinguishable from free for the user, and the advancing frontier of commercially refined knowledge products (see Fig.~\ref{fig1}). As open access becomes more the dominant mode of publication, the annulus cannot be solely positioned as a pathology nor as an information asymmetry imposed by rent-seeking commercial actors. Rather, it is a structural consequence of production friction in a system where the scholarly record was never designed for structured data consumption, and where the definition of what constitutes useful structured data is continuously evolving. The width of the annulus is, in a precise sense, a measure of the system's distance from perfect efficiency---and because perfect efficiency is logically impossible in a system where useful data types evolve faster than the production system can standardise them, the annulus is a permanent feature of the landscape.  In the AI era in which we now sit, we might expect AI to reduce production friction, but given the copyright complexities of text and data-mining and the challenges of understanding the algorithmic provenance of a piece of data so that it can be trusted \cite{Porter2026}, it is not clear that this assumption is well-founded.

This paper develops this argument in several steps. We begin by examining the cost structure of scholarly data production and the technological frictions that sustain the annulus (Sec.~\ref{sec:emh}). We then analyse the differentiated demand that gives the annulus its sectoral dimensions, develop a geometric interpretation of the annulus diagram that yields diagnostic measures including an openness ratio for each data type (Sec.~\ref{sec:differentiated-demand}), and present a formal welfare framework---analogous to the Nordhaus optimal patent life---for reasoning about the optimal annulus width (Sec.~\ref{sec:optimal-width}). We describe how artificial intelligence reshapes the annulus without eliminating it (Sec.~\ref{sec:ai-effect}), and examine structured in-paper metadata---CRediT author contributions, funding acknowledgements, AI disclosure---as frontier data types that illustrate the annulus lifecycle (Sec.~\ref{sec:integrity-case}). We then turn to the governance question: not whether the annulus should exist, but how its boundaries should be managed (Sec.~\ref{sec:governance}). We use the experience of Dimensions as an illustration of annulus dynamics in practice (Sec.~\ref{sec:dimensions-case}), and conclude with a set of open questions for the research community (Sec.~\ref{sec:research-agenda}).

\section{Production Friction and the Efficient Market Analogy}
\label{sec:emh}

The production of structured scholarly data is often discussed as though it were a problem that technology has essentially solved, leaving only political and institutional barriers to full openness. This view underestimates the depth and persistence of the frictions involved.

To see why, it is helpful to borrow a concept from financial economics. The efficient market hypothesis (EMH)~\cite{Fama1970} holds that, in a perfectly efficient market, all available information is reflected in prices and no actor can gain systematic advantage from private information. The EMH is not a description of how markets actually behave; it is a \emph{benchmark}---a theoretical limit against which the efficiency of real markets can be measured. Transaction costs, information asymmetries and regulatory frictions all cause real markets to deviate from the benchmark, and much of financial regulation is directed toward reducing these deviations.

We can apply the same logic to the scholarly data production system. In a perfectly efficient system---one in which all scholarly data were produced in a fully standardised, structured format at the point of creation, with universal identifiers, machine-readable metadata and complete provenance---there would be no need for downstream normalisation, enrichment or disambiguation. The information latent in the scholarly record would be fully expressed in its published form. The annulus would collapse to zero.

The actual scholarly data production system is, of course, very far from this benchmark. Scholarly data are produced by a highly heterogeneous publishing ecosystem that evolved over centuries without overall standardisation. Author names of individual researchers vary from publication to publication either through naturally arising inhomogenity or through more structured mechanisms such as the application of different publisher and journal house styles. Institutional affiliations are recorded inconsistently---a researcher may name their department but not their university, or use any of several variant forms (``University of Oxford'' versus ``Oxford University'' versus ``Dept. of Physics, Oxford''). Citation lists are formatted differently across thousands of journals. Funding acknowledgements may or may not include grant numbers, may name funders in full or use ambiguous abbreviations. These are not minor inconveniences; they represent deep structural frictions in the data production system that create a persistent demand for downstream normalisation~\cite{HookPorterHerzog2018}.

The annulus, by this account, is a direct measure of the system's production inefficiency. Its width for any given data type reflects the gap between how that data type is actually produced (with all its frictions, inconsistencies and omissions) and how it would need to be produced for downstream use to require no additional investment. Where the gap is large---as it historically has been for author disambiguation, institutional identification and citation linking---the annulus is wide, and there is economic space for actors who invest in closing the gap. Where the gap is small---as it is becoming for basic bibliographic metadata decorated with DOIs and deposited through Crossref---the annulus narrows and the data migrates toward the open core.

Three categories of production friction sustain the annulus even as technology advances:
\begin{enumerate}
\item \emph{Source heterogeneity.} The scholarly record comprises journal articles, conference proceedings, books, book chapters, preprints, patents, clinical trial registrations, policy documents, datasets, software, peer reviews~\cite{Teixeira2022}, seminars~\cite{Vincent2021,Gillett2021}, and more---each with different metadata conventions, identifier systems and quality standards. Structuring this heterogeneous landscape into a coherent, interlinked graph requires continuous investment in entity resolution, disambiguation and cross-referencing that scales with the diversity of the record rather than its volume alone~\cite{HookPorterHerzog2018,HookPorter2022}.

\item \emph{Quality decay.} Structured data does not remain accurate without maintenance. Researchers change institutions, journals change publishers, grants are reclassified, organisations merge or dissolve, and classification systems evolve. The cost of maintaining a living, accurate representation of the scholarly record is not a one-time investment but an ongoing operational expense---what Borgman~\cite{Borgman2015} has characterised as the continuous labour of knowledge infrastructure stewardship, and what Edwards et al.~\cite{Edwards2013} have documented as the invisible maintenance burden of knowledge infrastructures.

\item \emph{Frontier data types.} Even if every existing friction were resolved---if every author had an ORCID, every institution a ROR, every reference a DOI---the system would continue to generate new data types that begin their life in an unstructured, unstandardised state. Twenty years ago, nobody needed structured AI disclosure metadata because AI was not used at any significant scale in research. Ten years ago, few were thinking in terms of using structured research integrity signals at scale. The definition of what constitutes useful structured data is continuously expanding, and each new data type restarts the cycle of friction, investment, standardisation and eventual commoditisation.
\end{enumerate}

This last point is critical. It means the annulus is not merely a legacy of historical infrastructure debt that will be paid down over time. It is a \emph{permanent structural feature} of any knowledge system in which the definition of useful structured data evolves faster than the production system can standardise it. The annulus shrinks in specific places as standards mature---but it re-emerges at the frontier wherever new information needs outpace the system's native production capacity.

A deeper physical grounding for this claim is available and worth making explicit. In \emph{The Human Use of Human Beings}, Wiener framed organised systems---biological, mechanical, institutional, informational---as local enclaves that persist in defiance of the second law of thermodynamics only by continuously importing energy and exporting disorder to their surroundings~\cite{wiener1954human}. Structured information is, in this view, a form of negentropy; any system that maintains it must continuously do work against the universal tendency toward disorder. Left to themselves, messages accumulate noise, classifications decay, identifiers decouple from the entities they name, and standards fragment as they are locally reinterpreted. The three categories of production friction identified above are specific expressions of this tendency rather than peculiarities of the scholarly ecosystem: source heterogeneity is entropy in the generative process; quality decay is entropy accumulating in a previously ordered store; and the continuous emergence of frontier data types reflects the expansion of the informational phase space faster than any one-time investment can contain. The annulus, from this perspective, is the thermodynamic signature of the work being done against that disorder. Its width for any given data type is, in effect, a measure of the energetic cost of holding that region of the scholarly record in an analysable state; eliminating it would require an impossible condition---an information system in which organisation required no ongoing input. Maxwell's demon cannot get something for nothing, and neither can a metadata infrastructure: interventions that attempt to collapse the annulus do not remove the cost of maintaining order, they only reassign where that cost is paid. The annulus is the price of sustaining an anti-entropic enclave in an informational system that would, without continuous investment, drift toward unstructured noise.

The analogy with copyright is instructive here. Copyright creates a legally defined zone of exclusivity around creative works---a period during which the creator can recover their investment before the work enters the public domain. The width of this zone varies across domains (French moral rights for musical works extend further than Anglo-American economic copyright \cite{Peeler1999,Khan2008}) and reflects a social and value-driven judgment about the importance of protecting creative effort. PhD thesis embargoes in arts, humanities and social sciences represent a similar mechanism: the community has collectively agreed that a graduating student's interest in publishing a monograph justifies a temporary restriction on open access. In both cases, the width of the ``annulus'' is determined by a norm---legal, social or community-defined---that balances the public interest in access against the producer's interest in return on effort.

In the case of scholarly metadata, no such norm exists. The width of the annulus is determined not by statute or community agreement but dynamically, by the interaction of production cost and market demand. This is not necessarily a problem---a dynamically determined width may be more responsive to real changes in production economics than any fixed norm---but it does mean that the width is vulnerable to market power. The historically inflated annulus of the Web of Science / Scopus duopoly era~\cite{Lariviere2015,Mongeon2016} demonstrates what happens when the annulus width is sustained not by genuine production costs but by barriers to entry and institutional lock-in. The emergence of Dimensions~\cite{HookPorterHerzog2018}, OpenAlex~\cite{Priem2022} and expanded open infrastructure through Crossref has demonstrated that much of what was in the annulus could be produced more cheaply and made available more openly---compressing the annulus toward a width that more accurately reflects genuine production friction.

It is worth noting that the annulus width for any given data type reflects not only technical production friction but also \emph{legal and contractual friction}. Licensing restrictions on full-text access, for example, widen the annulus independently of the technical cost of mining. Even if AI were to reduce the computational cost of extracting structured metadata from full-text articles to near zero, the inability to access full text at scale---because of copyright restrictions, publisher licence terms, or paywalled content---keeps the outer boundary (the structuring frontier) further from the centre than the technical economics alone would dictate. A complete analysis of annulus width for any given data type must therefore decompose it into its technical component (the genuine cost of structuring) and its legal component (the artificial friction created by access restrictions). The governance prescriptions for reducing these two types of friction are quite different: technical friction is addressed by standards adoption and technological investment; legal friction is addressed by licensing reform, collective disclosure agreements and policy intervention.

\section{Differentiated Demand and the Sectoral Annulus}
\label{sec:differentiated-demand}

The annulus model as described so far treats the demand side as undifferentiated. In practice, the demand for structured scholarly data is radically heterogeneous, and this heterogeneity gives the annulus sectoral dimensions that move at different speeds and have different governance implications.

We identify three demand drivers that create distinct quality requirements, each of which shapes the width of the annulus in a different part of the knowledge spectrum.

\subsection{Competitive strategy and information asymmetry}
\label{sec:competitive-drive}

Research institutions compete for funding, talent and reputation. This competition is not a market distortion but a deliberate feature of research systems designed to allocate scarce resources to the most productive groups~\cite{Dasgupta1994}. Competition creates demand for information asymmetries: institutions that can identify emerging research strengths more rapidly than and prior to their competitors, benchmark their performance more precisely, or anticipate funder priorities more accurately gain a structural advantage. This mechanism leads to the creation of centres of excellence and a varied landscape with specialisms. This dynamic is analogous to the knowledge spillovers that Jaffe~\cite{Jaffe1986} demonstrated in industrial R\&D: structured data about who is doing what research, is the mechanism through which competitive intelligence flows, and actors who can access and interpret that data more effectively gain a systematic advantage.

This demand is not served by raw open data. It is served by refined, contextualised analytical products that transform the scholarly record into strategic intelligence. University research offices invest in benchmarking tools not because they object to openness but because their competitive environment rewards the ability to extract actionable insight faster and more accurately than other actors. The data they need is not merely \emph{accessible} in the FAIR sense~\cite{Wilkinson2016}---it must be Findable, Interoperable and Reusable in a specific institutional context, which currently requires curation and enrichment beyond what open provision typically delivers.

\subsection{Research translation and alignment data}
\label{sec:translation}

An increasingly important driver is the emerging ecosystem of research translation partnerships between academic and commercial organisations. The translation of basic research into application---what Stokes~\cite{Stokes1997} mapped as ``use-inspired basic research''---requires both parties to understand the alignment of their respective capabilities. In what Adams~\cite{Adams2013} has characterised as the ``fourth age of research,'' where the dominant mode of production is international collaboration, the data infrastructure required to support translation must be correspondingly global and interlinked. Which research groups are working on problems relevant to a company's R\&D pipeline? Which institutions have the expertise and infrastructure to support a translational challenge? These questions require high-quality data on research capabilities that goes beyond publication counts: entity-resolved researcher profiles, institutionally contextualised output data, patent-to-publication linkage, clinical trial mapping and funding flow analysis---all at a quality sufficient to support investment decisions~\cite{HookPorter2022}.

\subsection{Corporate research and domain-specific refinement}
\label{sec:corporate-demand}

Research-intensive industries---biomedical and pharmaceutical most prominently, but also automotive, engineering, agrichemical, petrochemical and financial, to name just a few---conduct substantial research programmes and depend on the scholarly record for competitive intelligence, prior art analysis and strategic planning. Their data needs differ from those of academic users in important respects: they require data at a granularity and domain specificity that reflects their particular R\&D portfolios; they are willing to pay for quality because the decisions that depend on it have direct financial consequences; and they often require integration across scholarly and non-scholarly sources---patents, regulatory filings, clinical trial registrations, market data---that falls outside the scope of what open scholarly infrastructure was designed to provide. Even at the national policy level, analyses demonstrating that research impact is driven primarily by international collaboration rather than domestic performance~\cite{AdamsSzomszor2024} require data structured and refined to a quality that supports robust interpretation~\cite{Szomszor2020}---a quality level that raw open metadata does not yet consistently deliver.

\subsection{Reading the annulus geometry}
\label{sec:geometry}

\begin{figure*}[t]
\centering
\includegraphics[width=\textwidth]{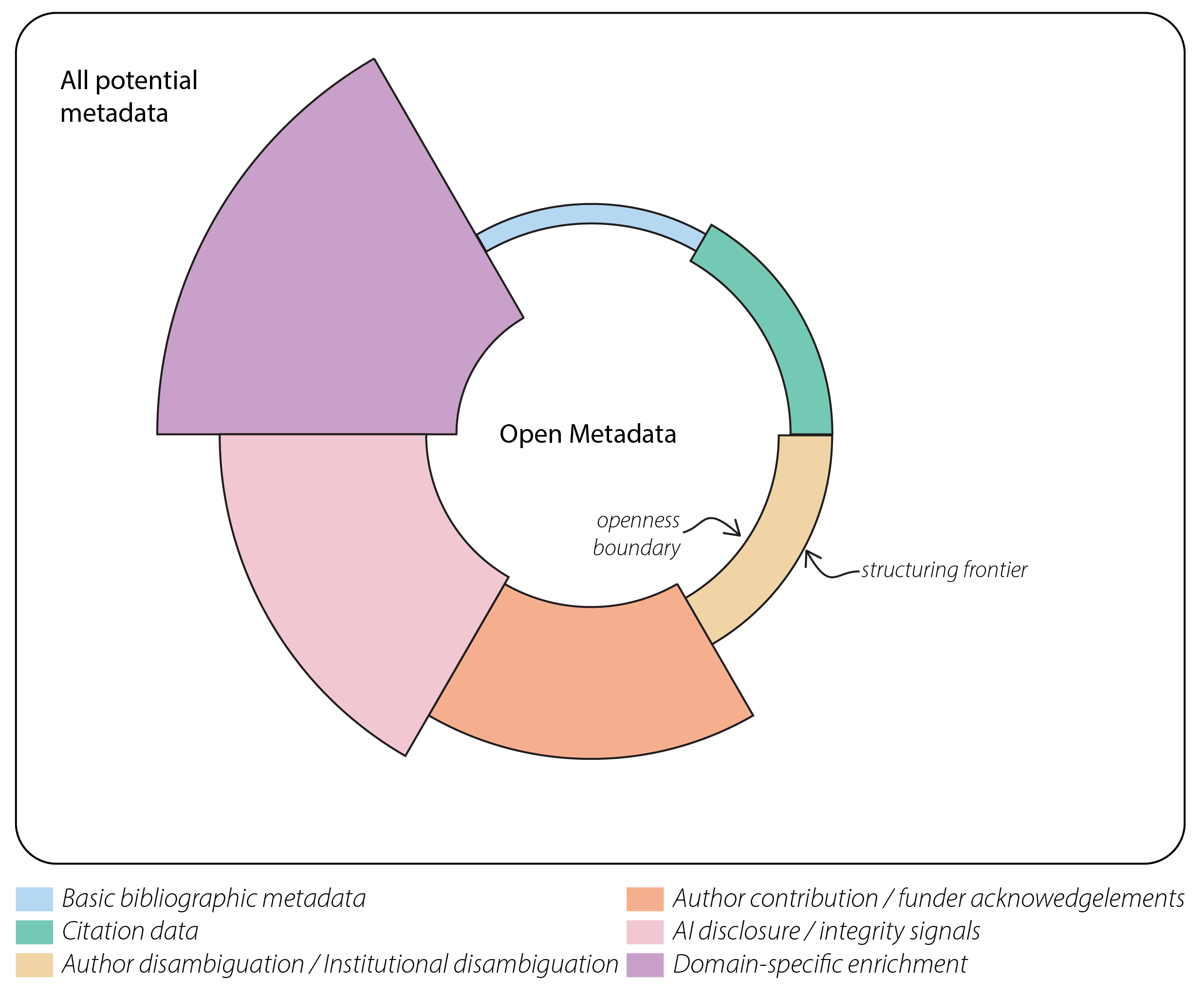}
\caption{\label{fig:sectoral}\small\textbf{Sectoral Dimensions of the Annulus.} The concentric-circle diagram from Figure~\ref{fig1}, but with the annulus divided into radial sectors of different widths and different radial positions, each labelled with a data type. Critically, the inner and outer arc of each sector are positioned independently. The radial distance from the centre represents the total structuring investment and ecosystem development for that data type: segments further from the centre indicate more mature, more developed, and more extensively structured domains. The thickness of each segment represents the commercial opportunity---the gap between what is openly available and what has been structured at the frontier. }
\end{figure*}

The sectoral annulus model introduced above can be made more precise by attending carefully to the geometry of the diagram (Fig.~\ref{fig:sectoral}). Each data-type segment has two independent visual properties: its \emph{radial position} (how far from the centre both the inner and outer arcs sit) and its \emph{thickness} (the gap between the inner and outer arcs). Together, these encode a surprisingly rich set of inferences about the state of a data type in the knowledge ecosystem.

The \emph{radial position} of a segment captures the total structuring investment and ecosystem development for that data type. If both arcs are close to the centre, the data type is either embryonic (nobody has invested much in structuring it because demand has not yet materialised) or inherently limited in scope (the information content is bounded and does not require deep development). If both arcs are far from the centre, the data type has a well-developed ecosystem with significant structuring effort and mature systems. Radial position is not, however, purely about demand: a data type may be close to the centre either because there is little interest in it or because the cost of mining it is prohibitively high despite significant latent demand. The distinction matters for governance: the first case requires no intervention, while the second may require public investment or standards development to unlock the latent value.

The \emph{thickness} of a segment represents the commercial opportunity---the gap between what is openly available and what has been structured at the frontier. A thick segment indicates a large space in which commercial refinement can operate. A thin segment indicates that open provision is close to the frontier, leaving little room for commercial differentiation.  

These two properties interact to create four characteristic configurations:
\begin{enumerate}
\item A segment that is \emph{thin and close to the centre} represents a data type that is low-cost to produce, for which adequate systems are in place, and where little innovation is required. Structured peer review metadata might currently fall into this category---some data exists, but neither demand nor production investment is deep.

\item A segment that is \emph{thin and far from the centre} represents the success case: mature systems, good standards, and open provision that has caught up with the frontier. Basic bibliographic metadata with DOIs is approaching this state. The annulus has been compressed by standards adoption and collective disclosure through Crossref.

\item A segment that is \emph{thick and close to the centre} represents either gatekeeping (significant structuring has occurred but little is openly available) or an embryonic field where the data are expensive and difficult to mine.  The legacy Web of Science model exemplified a gatekeeping pattern: the inner arc was very close to the centre (almost nothing open) while the outer arc was moderately distant (substantial commercial product). This makes sense during the initial development of a new data source such as the Web of Science. In the 1950s the data were not diverse, it was hard to mine and hence a homogeneous annulus would be an appropriate representation similar to Figure~\ref{fig1}). The openness boundary can be shrunk to the centre with the annulus becoming a fully filled circle as there was essentially no open data---the cost of production precluded that option at that time---the data were excessively expensive to mine. Indeed, so expensive was it to mine data that it was only Garfield's critical realisation was that citations were Pareto distributed \cite{Bradford1934,Garfield1980} that made it financially tractable to construct an index \cite{Mills2024,Wouters1999}. For more recent emerging data types like AI disclosure, modern technologies automatically set the boundaries differently---the thickness reflects tends to be closer to the market expectation of the production cost.

\item A segment that is \emph{thick and far from the centre} indicates a data type with extensive development at the frontier and a substantial commercial zone, but also a meaningful open core. Domain-specific enrichment---pharmaceutical patent-to-publication linkage, translational alignment data, institutional analytics---sits here. The thickness reflects the genuine cost of refinement at the quality levels that demanding users require.
\end{enumerate}

Several additional inferences follow from this geometry. First, the ratio of the inner radius to the outer radius for each segment yields a natural \emph{openness ratio}---the proportion of total structuring effort that is openly available. A ratio approaching one indicates near-complete openness for a mature data type. A ratio close to zero indicates that most structured data remains behind a commercial boundary. This ratio is a useful diagnostic for governance: one might argue that for data types essential to basic research management, the openness ratio should be above some threshold, even if the absolute frontier extends further for specialised users.

Second, the diagram should be understood as a snapshot of a dynamic system. The trajectory of each segment is as important as its current position. A segment in which both arcs are moving outward indicates a dynamic, maturing data type---new structuring is occurring and new data is becoming open. A segment in which the outer arc moves outward while the inner arc remains stationary represents increasing commercial development without corresponding openness gains---a pattern that should raise governance concerns. A segment in which the inner arc is catching up with the outer represents active commoditisation. And a segment in which neither arc moves is stagnant.

Third, there is a structural constraint that the diagram makes visible: the inner boundary (or openness boundary) can never move outward faster than the outer boundary (the structuring frontier) in the long run, because data cannot be made openly available until it has been structured. Open provision is bounded above by total structuring effort. This has a crucial implication for the argument of this paper. In domains where no one is investing in frontier structuring---because there is no commercial incentive and no public funding---the inner boundary stalls too. The annulus does not merely describe the commercial gap; it describes the \emph{investment incentive}. An annulus that is too thin removes the economic incentive for frontier structuring, which in turn slows the outward movement of the inner boundary. This is the strongest version of the argument that the annulus is functional: without sufficient annulus width, the entire system---including the open core---advances more slowly.

Fourth, there is an important category of cases in which the inner arc is further out than market forces alone would produce. This represents \emph{deliberate intervention}---a mandate, a philanthropic investment, or a strategic commercial decision to make data open beyond what the natural economics would deliver. The Dimensions free tier and BigQuery access~\cite{Herzog2020,HookPorter2021} represent strategic commercial decisions to push the inner boundary outward. OpenAlex~\cite{Priem2022} represents philanthropic funding achieving a similar effect. The Barcelona Declaration~\cite{BarcelonaDeclaration2024} is, in effect, an attempt to create coordinated pressure to push the inner boundary outward across many data types simultaneously. The inner boundary is not solely determined by production cost---it is also shaped by values, strategy and governance choices.

\subsection{The equity dimension}
\label{sec:equity}

A consideration that cuts across the three demand drivers and the geometric analysis above concerns the equity implications of how the annulus is structured. In a world where AI can be used to enhance and enrich base-level data, institutions with computational resources, technical talent and data science capacity can run their own enhancement pipelines---resolving affiliations, classifying output, identifying collaboration opportunities. Institutions in under-resourced settings cannot. The data is nominally open, but the capacity to extract value from it is profoundly unequal.

This creates a strong case for centralised provision at accessible cost. If each institution must independently enhance its own data, the gap between rich and poor institutions widens. If intermediaries---whether commercial or community~\cite{Herzog2020}---produce enhanced, trustworthy data that everyone can agree is a source of ``truth'' for certain evaluative activities at a consistent quality standard, the baseline rises for everyone. The centralised model is not just more computationally efficient; it is more equitable, because it sets a floor below which no institution needs to fall~\cite{Czepan2024}. Lane~\cite{Lane2020} has articulated this principle forcefully in the context of public data systems: Data infrastructure should be designed so that access does not depend on insider knowledge or privileged networks, because ``unless a researcher is able to tap into a network of cognoscenti, they would not know about the data, or not know how to use it''~\cite{LanePotok2024}. The same logic applies to scholarly metadata: the open core must be not merely technically accessible but practically usable by institutions at all resource levels. Adams, Gurney, Hook and Leydesdorff~\cite{AdamsGurneyHook2014} have demonstrated what structured data can reveal about collaboration patterns in Africa---analyses that would be impossible without the kind of institutionally resolved, openly available metadata that the inner circle of the annulus is designed to provide.  More recently Pinfield~\cite{pinfield2024} argues that nominal openness isn't sufficient when the capacity to extract value from open data is unequally distributed.

The governance implication is that public investment should focus on ensuring that the baseline of structured data available to all institutions is as high as possible---not on eliminating the annulus entirely. The annulus above the baseline serves differentiated demand from users who can afford to pay for frontier refinement. The baseline---the inner circle---serves the equity function of ensuring that under-resourced institutions are not excluded from the structured data they need for competent research management and strategic decision-making.  In some sense, this constitutes a modern form of institution building---in this case not research-performing institutions but rather social institutions that define norms and expectations of data, its quality and its uses.

\section{Toward an Optimal Annulus Width}
\label{sec:optimal-width}
The geometric framework developed in Sec.~\ref{sec:geometry} makes the annulus visually tractable. The natural next question is whether it can be made analytically tractable: Is there a principled way to determine what the annulus width \emph{should} be for a given data type, rather than merely observing what it \emph{is}?

The question has a structural analogue in the economics of intellectual property. Nordhaus~\cite{Nordhaus1969} posed the same question for patents: given that a period of monopoly exclusivity is needed to incentivise invention, what is the optimal length of that period? His answer---a formal optimisation trading off the deadweight loss of monopoly pricing against the incentive to invest in R\&D---established a framework that has been extended by Klemperer~\cite{Klemperer1990}, Gallini~\cite{Gallini1992} and Scotchmer~\cite{Scotchmer2004}, and that remains foundational in innovation economics. We argue that the annulus presents a structurally homologous problem, and that a similar framework can be developed for scholarly knowledge infrastructure.

\subsection{A welfare framework}
\label{sec:welfare-framework}
Consider a single segment of the annulus at a given point in time. Let $r_i$ denote the inner radius (the openness boundary) and $r_o$ the outer radius (the structuring frontier). The annulus width is $w = r_o - r_i$ and the openness ratio is $\rho = r_i / r_o$.

It is helpful to define the value and cost functions in an order that parallels how data enter the system: structuring first, and open provision as an overlay on structured data. We therefore begin with the total value and cost of structuring, and then introduce separately the additional welfare and different cost profile associated with making part of the structured data openly available.

\textit{Value of structuring.} Let $V(r_o)$ denote the total value, to all users, of having data structured up to frontier level $r_o$, evaluated at whatever access terms apply (paywalled, tiered or free). $V$ is increasing and concave in $r_o$: deeper structuring yields diminishing returns as progressively more specialised data types are brought into the structured domain.

\textit{Cost of structuring.} Let $C(r_o)$ denote the full cost of producing and sustaining structure up to level $r_o$---entity resolution, disambiguation, classification, and the ongoing stewardship required to keep the structured representation current as the underlying record evolves~\cite{Edwards2013,Borgman2015}. $C$ is increasing and convex: the easy structuring tasks (for example, applying DOIs, resolving common institutional names) are accomplished first, and each additional unit of frontier structuring requires more domain expertise, more validation and continuing maintenance. We do not separate production from maintenance because they are incurred jointly and scale together with $r_o$.

\textit{Openness premium.} Let $B(r_i)$ denote the \emph{additional} welfare gain from making the range $[0, r_i]$ openly available rather than providing it on restricted terms---the equity, access and knowledge-spillover premium that openness is incremental to the value captured in $V$. $B$ is increasing and concave in $r_i$: the first units of open provision (basic bibliographic metadata, core identifiers) have enormous marginal value because they set a floor for all institutions, while later units have diminishing marginal impact. This is where the equity argument of Sec.~\ref{sec:equity} enters formally---the social weight on the early units of $B$ is high because they serve under-resourced institutions that would otherwise be excluded~\cite{Stiglitz2000}.

\textit{Cost of the openness overlay.} Let $M(r_i)$ denote the additional cost of running the openness overlay on the range $[0, r_i]$: data standardisation and governance, free distribution at scale, community oversight, and the persistence and preservation guarantees that openness requires. $M$ is increasing in $r_i$ and is subject to external pressures that may raise it exogenously---the AI-harvesting cost inflation documented by Crossref~\cite{Hendricks2023} enters through $M$ rather than $C$.

Under this decomposition, paywalled data inside the annulus incurs $C$ but not $M$; open-core data incurs both. The social welfare function is then
\begin{equation}
W \;=\; V(r_o) + B(r_i) - C(r_o) - M(r_i).
\label{eq:welfare}
\end{equation}

This is subject to a sustainability constraint: the system must be financially viable. In general, revenue depends on both boundaries, since deeper frontier refinement commands higher marginal revenue per unit of width than shallow refinement of commoditised data. The appropriate object is therefore $R(r_i, r_o)$. Within a single segment of the annulus, where the radial range is narrow, revenue is locally well approximated as a function of width alone, and we write $R(w)$ with the understanding that this is a local reduction. We return to this point in Sec.~\ref{sec:patent-theory}. Revenue plus any public subsidy $S$ must cover the costs:
\begin{equation}
R(w) + S \;\geq\; C(r_o) + M(r_i).
\label{eq:constraint}
\end{equation}

The constrained optimisation yields first-order conditions that characterise the optimal boundaries. For the inner boundary,
\begin{equation}
B'(r_i) \;=\; (1+\lambda)\,\bigl[M'(r_i) + R'(w)\bigr],
\label{eq:foc-inner}
\end{equation}
and for the outer boundary,
\begin{equation}
V'(r_o) \;=\; (1+\lambda)\,\bigl[C'(r_o) - R'(w)\bigr],
\label{eq:foc-outer}
\end{equation}
where $\lambda \geq 0$ is the shadow price of the sustainability constraint and a dash implies the derivative of the function with respect to its natural variable, thus $B'(r_i)={\rm d}B/{\rm d}r_i$. $\lambda$ measures the social welfare gained from relaxing the constraint by one unit---for example, through an incremental unit of public subsidy. When revenue plus subsidy more than covers costs, the constraint does not bind and $\lambda = 0$; the first-order conditions reduce to the unconstrained conditions $B'(r_i) = M'(r_i) + R'(w)$ and $V'(r_o) = C'(r_o) - R'(w)$. When the constraint binds tightly, $\lambda$ is large and the annulus must do more of the work of sustaining the system.

The first-order conditions have a direct intuitive reading. Equation~\eqref{eq:foc-inner} states that the marginal social value of expanding the open core, $B'(r_i)$, should equal the marginal cost of doing so---directly, through the incremental cost of open provision, $M'(r_i)$, and indirectly, through the revenue forgone by narrowing the annulus, $R'(w)$. The weight $(1+\lambda)$ reflects the fact that, when money is tight, costs and forgone revenues count more heavily against social welfare. Equation~\eqref{eq:foc-outer} is the symmetric condition for the outer boundary: the marginal value of frontier structuring, $V'(r_o)$, should equal the marginal structuring cost, $C'(r_o)$, less the marginal revenue gained from widening the annulus, $R'(w)$, again weighted by the tightness of the financial constraint. Readers seeking a textbook treatment of the underlying constrained-welfare machinery will find the most relevant material in Stiglitz~\cite{Stiglitz2000} and, at greater technical depth, in Atkinson and Stiglitz~\cite{atkinson1980lectures}.

\begin{figure*}[t]
\centering
\includegraphics[width=\textwidth]{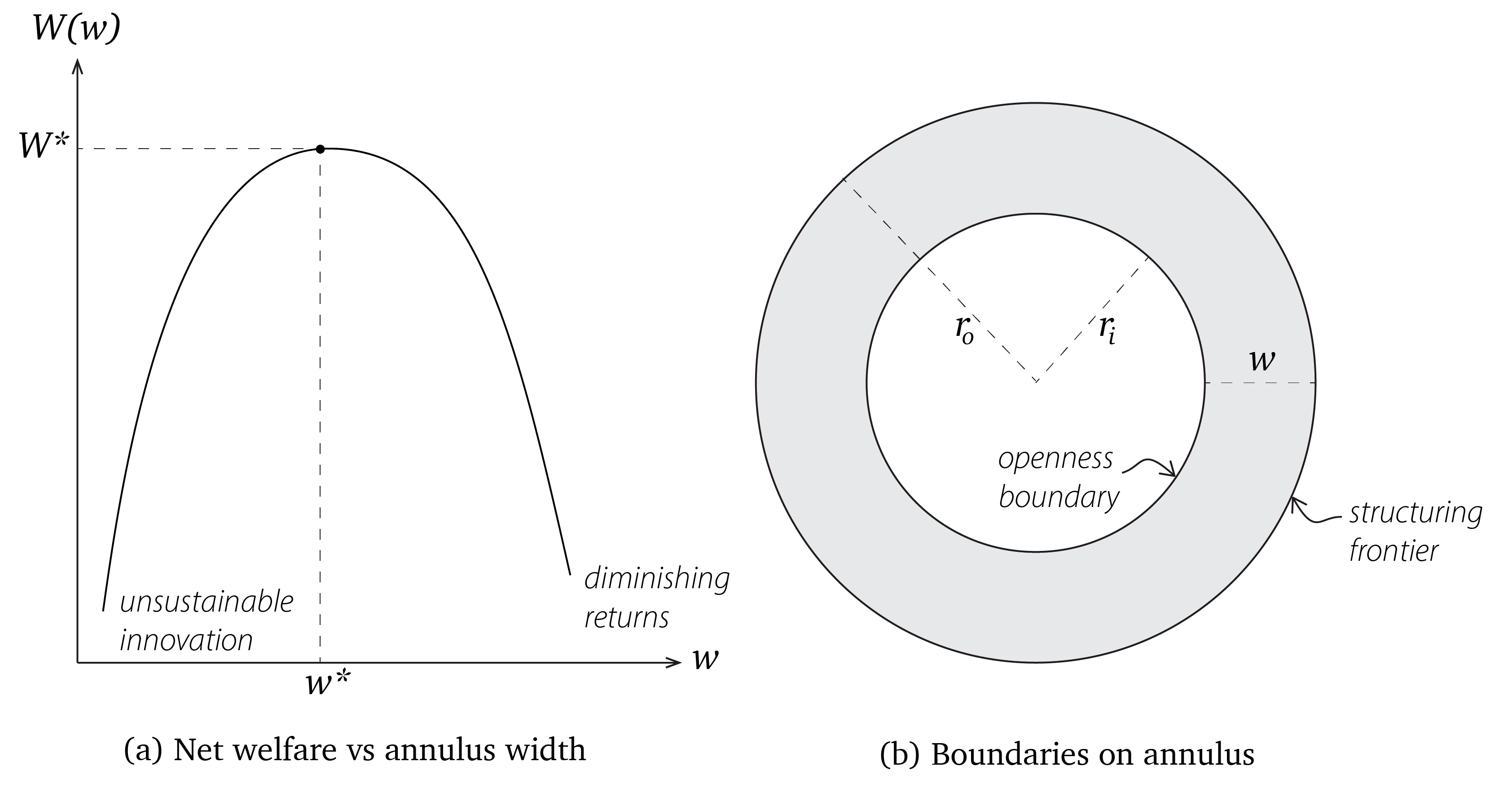}
\caption{\textbf{A welfare-theoretic view of the optimal annulus width.} Panel~(a) plots net welfare $W$ as a function of annulus width $w$, holding the outer boundary at its optimal level. The inverted-U is asymmetric: the left shoulder reflects a discrete feasibility collapse as the sustainability constraint, Eq.~\eqref{eq:constraint}, ceases to hold and frontier structuring becomes unsustainable; the right shoulder reflects the continuous opportunity cost of forgoing units of $B(r_i)$. Panel~(b) shows the boundaries $r_i$ and $r_o$ and the width $w$.}
\label{fig:welfare}
\end{figure*}

Figure~\ref{fig:welfare} explores these conditions pictorially. Panel~(a) plots net welfare $W$ as a function of annulus width $w$, holding the outer boundary at its optimal level: the curve is an inverted-U with maximum at $w^*$. As $w \to 0$ the sustainability constraint binds and frontier investment becomes unsustainable; as $w$ grows large the social benefit of openness is progressively forgone. The asymmetry of the curve is deliberate: the left shoulder reflects a discrete feasibility collapse (the constraint ceases to hold), while the right shoulder reflects a continuous opportunity cost (forgone units of $B(r_i)$). Panel~(b) shows the boundaries $r_i$ and $r_o$ and the width $w$ of the annulus.

\subsection{Qualitative predictions}
\label{sec:predictions}

Even without solving the optimisation for specific functional forms, the framework yields several qualitative predictions that are, in principle, testable.

\textbf{Prediction 1: the optimal annulus width is wider where structuring costs are higher.} When the marginal cost (i.e. the cost of obtaining an additional data point in a given set) of structuring data $C'(r_o)$ is large, the sustainability constraint binds more tightly (higher $\lambda$), and more revenue from the annulus is needed to sustain frontier investment. This explains why the annulus persists for data types that require deep domain-specific enrichment (pharmaceutical patent linkage, translational alignment data) while narrowing for data types where AI has reduced structuring costs.

\textbf{Prediction 2: the optimal width is narrower where the social benefit of openness is steep.} When the marginal welfare gain (i.e. the welfare associated with making open an additional data point in a given data set) $B'(r_i)$ is large---for data types essential to basic research management, where equity concerns dominate---the first-order condition~\eqref{eq:foc-inner} pushes $r_i$ outward, narrowing the annulus. This formalises the intuition that governance should prioritise high openness ratios for foundational data types.

\textbf{Prediction 3: the optimal width shrinks as technology reduces structuring costs.} As AI and standards adoption reduce $C(r_o)$, the sustainability constraint loosens (lower $\lambda$), and the inner boundary can move outward. This is the formal version of the claim that technology compresses the annulus---but it also predicts that the compression is modulated by the shadow price, not automatic.  There are a variety of assumptions implicit in this.  Two of the most significant are that: i) the externalities associated with data labelling and the environmental impact of AI are negligible; ii) that algorithmic data improvement finds a systematic structure that allows context of data improvements to travel with metadata in a transparent manner.

\textbf{Prediction 4: public subsidy is a substitute for annulus width.} A larger $S$ relaxes the sustainability constraint, allowing a thinner annulus at the same level of frontier investment. The Entrepreneurial State model~\cite{Mazzucato2013} corresponds to the limiting case $S \to S^{*}$ where subsidy fully replaces annulus revenue and $w \to 0$. The functional annulus model corresponds to moderate $S$ and moderate $w$. The framework makes explicit what each model asks of the public purse.  The risk of this model is that innovation may lack an appropriate efficiency moderator.

\textbf{Prediction 5: an annulus that is too thin slows the entire system.} This follows from the structural constraint that the inner boundary cannot advance faster than the outer boundary: $r_i \leq r_o$ at all times. If the annulus is compressed below the level at which frontier investment is sustainable ($R(w) + S < C(r_o) + M(r_i)$), the outer boundary stalls---and with it, the inner boundary. Open provision is bounded above by frontier structuring. This is the formal expression of the argument that the annulus is functional: eliminating it does not merely reduce commercial revenue but reduces the rate at which the open core can expand.

\subsection{Relation to patent theory and limitations}
\label{sec:patent-theory}

The framework is structurally homologous to the Nordhaus optimal patent life derivation~\cite{Nordhaus1969}, but differs in two important respects. First, in the patent case, the deadweight loss arises from monopoly pricing during the exclusivity period---a dynamic that Jaffe and Lerner~\cite{JaffeLerner2004} have documented can lead to patent systems in which exclusivity periods bear little relationship to the investment required. In the annulus case, the analogous loss arises from restricted data access: institutions that cannot afford refined data make worse decisions than they would with full access, and the research ecosystem as a whole operates below its potential~\cite{Arrow1962}. Second, patent life is uniform across all inventions within a jurisdiction. The optimal annulus width is a function of the data type's characteristics---its structuring cost profile, its demand heterogeneity, and the social weight placed on open access to it. The framework does not yield a single optimal width but an optimal width function $w^{*}(C', B', V', \lambda)$ that varies across segments of the annulus diagram. This is more complex than the patent case but also more realistic, since different data types in the scholarly ecosystem clearly operate under different economic conditions.

The framework has important limitations. First, the functional forms of $B$, $V$, $C$, $M$ and $R$ are not known empirically for any data type in the scholarly ecosystem; their estimation is itself a research programme (see Sec.~\ref{sec:research-agenda}). Second, the shadow price $\lambda$ is not directly observable; proxies would need to be developed---for example, the ratio of unmet demand for open data to available public funding could serve as a coarse indicator of when the constraint binds tightly. Third, the reduction from $R(r_i, r_o)$ to $R(w)$ is a local approximation, valid within a single segment of the annulus where the radial range is narrow. Across segments, and at very different radial positions, revenue depends on the absolute level of structuring as well as on its differential from the open core, and the general form $R(r_i, r_o)$ is appropriate; the qualitative predictions of Sec.~\ref{sec:predictions} are robust to this generalisation because they depend on signs of first and second derivatives rather than on the specific reduction to width. Fourth, the model is static: it characterises the optimal annulus at a point in time rather than the optimal trajectory of both boundaries. A dynamic extension---casting the problem as an optimal control problem with equations of motion for $r_i(t)$ and $r_o(t)$, driven by technological change and standards adoption---would be a natural next step, and would connect to the broader literature on the optimal timing of technology diffusion.

We note these limitations not to undermine the framework but to identify the empirical and theoretical work required to make it operational. Even in its current form, the framework provides something that the governance conversation around open research information has lacked: a formal language for reasoning about the trade-offs involved in setting the boundary, and a set of testable predictions about how the annulus should respond to changes in technology, demand and policy.

\section{AI and the Annulus}
\label{sec:ai-effect}

Artificial intelligence has transformed the economics of the annulus, but the nature of the transformation is frequently mischaracterised. The common narrative holds that AI will collapse the annulus entirely by automating metadata extraction, entity resolution and classification to near-zero cost. The framework developed in Sec.~\ref{sec:optimal-width} shows why this is too simple: AI reduces $C(r_o)$, which loosens the sustainability constraint and allows the inner boundary to move outward, but it does not drive $C$ to zero and it simultaneously affects $M$, $V$ and the demand structure. Three specific dynamics warrant attention.

\subsection{Frontier acceleration and cost reduction}

AI techniques---automated metadata extraction, entity resolution, semantic classification, citation parsing---have lowered the cost and increased the speed of basic data structuring. Tasks that previously required teams of manual curators can now be accomplished computationally at a fraction of the cost. This is unambiguously positive: it pushes the inner boundary of the annulus outward, expanding the volume of data that can be provided openly, with ``appropriate provenance''.

Additionally, the distribution of benefit is uneven. AI is most powerful when applied to large, well-organised data collections. Actors who already hold substantial structured data assets gain disproportionate advantage because their existing assets provide training data, ground truth and computational context~\cite{ShapiroVarian1999}.

\subsection{Quality threshold elevation}

A less-discussed effect of AI is that it \emph{raises the quality threshold} at which data refinement has commercial value. As basic structuring becomes commoditised through AI automation, the annulus does not disappear---it migrates to higher-order refinement tasks. The commercially valuable frontier moves from ``can you structure this data at all?'' to ``can you structure it at a quality level that supports investment decisions, regulatory submissions or competitive strategy?''

This dynamic is familiar from other technology-intensive industries. In financial data, the commoditisation of basic market data did not eliminate the market for refined analytics; it created a new premium tier of algorithmic intelligence. In geospatial data, the availability of free satellite imagery did not eliminate demand for domain-specific analysis; it shifted the value frontier from data acquisition to data interpretation. The same pattern is visible in scholarly data.

\subsection{Systemic risks of unprovenanced AI-derived metadata}

AI also introduces a systemic quality risk that has received insufficient attention. When multiple actors independently use AI to enhance metadata without tracking the provenance and processing history of their enhancements, the result is not merely duplicated effort---it is a potential quality degradation. An AI-derived institutional affiliation that is contributed back into a shared system without its processing history becomes, for the next consumer, an apparently authoritative data point whose reliability cannot be assessed. If that consumer's AI then builds on it, errors compound. Porter~\cite{Porter2024,Porter2026} has articulated this risk clearly: AI enhancement of metadata without understanding or context can lead to poorer quality data, as downstream users do not understand the full processing and provenance of a piece of metadata that has been contributed to a centralised system without the details of its history.

This creates a system-level efficiency argument for centralised structured data intermediaries---whether open or commercial---that goes beyond the usual access debate. The intermediary exists not to restrict access but because centralised normalisation with provenance tracking is more efficient and more reliable than distributed rederivation without it. In an era of concern about computational carbon footprints, the duplication cost of many independent AIs repeatedly rederiving the same structured data---rather than consuming it from a maintained, provenanced central store---is itself a consideration.

\section{Structured In-Paper Metadata: A Frontier Case Study}
\label{sec:integrity-case}

To illustrate the annulus lifecycle concretely, we examine a set of data types that are currently at different points on the journey from unstructured free text to standardised, openly available metadata. These are not data \emph{about} papers derived by external analysis (retraction databases, image manipulation detection) but data produced \emph{by authors as part of the publication process} that are not yet systematically structured into the formal scholarly record.

\emph{Funding acknowledgements.} Information the funding of a piece of research is present in the vast majority of papers but is structured in wildly inconsistent ways. Some authors include full funder names and grant numbers; others use ambiguous abbreviations; still others include only a brief narrative acknowledgement. Crossref's funder registry has begun to standardise this, and publishers increasingly request structured funding information---but the legacy record and the inconsistency of current practice mean that producing reliable, analysis-ready funding data at scale requires significant investment in natural language processing and entity resolution.

\emph{CRediT author contribution statements.} The CRediT taxonomy~\cite{Allen2014,Brand2015,McNutt2018} provides a standardised vocabulary for describing author contributions (conceptualisation, methodology, writing, supervision, etc.).  CRediT was designed to address the inadequacy of ordered author lists as a mechanism for attribution and credit~\cite{Allen2019}. Adoption is growing but uneven. A recent retrospective analysis found that as of 2024, only 22.5\% of original research articles with available full text in Dimensions included CRediT role information, with significant variation across publishers, disciplines and countries~\cite{Allen2025}. Where CRediT statements are present, they are not always machine-readable; where they are machine-readable, they are not always deposited in Crossref metadata. The Dimensions team, for example, has invested in the creation and curation of AI models that identify author contribution statements across the literature---work that operates at accuracy levels that still require improvement and hence further investment~\cite{Hook2024blog}. These data would be of significant value to the evaluation community and to anyone involved in tenure and promotion processes, but no universally accepted structured data format yet makes them widely available.

\emph{Data availability statements.} Many journals now require authors to declare whether and where their research data are available. These statements are typically free-text, with no standard vocabulary or structure. Extracting structured information about data availability at scale, particularly distinguishing between ``data available on request,'' ``data deposited in [specific repository],'' and ``no data were generated'' requires significant processing.

\emph{AI disclosure.} The most recent addition to this category, AI usage disclosure is currently required by a growing number of journals but in entirely unstandardised forms. The disclosure may appear in the methods section, the acknowledgements, a dedicated statement, or nowhere at all. There is no agreed vocabulary for describing what role AI played (writing assistance, data analysis, code generation, image creation). This represents a data type at the very outer edge of the annulus: enormously valuable for understanding the evolution of research practice, but currently extractable only through expensive, error-prone natural language processing.

Each of these data types follows a recognisable lifecycle. They begin as unstructured free text within papers (wide annulus, high production cost for anyone wanting to use them at scale). They move through a standardisation phase in which a taxonomy or identifier system is developed (CRediT, Crossref funder registry). They pass through a phase of uneven adoption in which the standard exists but is not universally applied. And eventually---though this has not yet happened for most of these data types---they become part of the baseline structured record that publishers produce natively, when the annulus for that data type shrinks toward zero.

The progression is real but slow, and it is driven by the same forces that the EMH analogy identifies: standards adoption that reduces production friction, collective disclosure agreements that expand the open baseline, and technological advancement that lowers the cost of extraction and normalisation. Cross-publisher efforts to standardise research data policies~\cite{Hrynaszkiewicz2017} and emerging initiatives to rethink publication models around open science principles~\cite{Kiermer2025} represent governance interventions that accelerate this progression by reducing production friction at source. Research linking publications to deposited data has demonstrated measurable citation advantages~\cite{Colavizza2020}, providing empirical evidence that structured metadata creates value---and hence that the investment in structuring is justified. It is worth noting that if originators---authors and publishers---had perfectly structured these data at source, with standard taxonomies and machine-readable formats, the normalisation annulus for these data types would not exist. The annulus is, in this precise sense, a consequence of historical and ongoing production inefficiency. As the publishing system modernises, and unique identifiers and structured data standards become more universal, the cost of data production and maintenance for these established data types will decrease. But new data types that the ecosystem will then want to consume---signals that we cannot yet anticipate---will create new frontiers where the same dynamic plays out again.

\section{Governance: Managing the Boundary}
\label{sec:governance}

If the annulus is a permanent structural feature of the knowledge ecosystem rather than a pathology to be eliminated, then the governance question changes. It is no longer ``how do we make everything open?'' but ``how do we ensure the annulus is the right width?'' Too thin, and there is insufficient economic incentive for the frontier data production that serves differentiated demand and drives innovation in data quality. Too thick, and the research ecosystem is poorly served---locked into paying for data above its actual value.

\subsection{The Entrepreneurial State and its limits}

The most developed case for public intervention in knowledge infrastructure is Mazzucato's Entrepreneurial State~\cite{Mazzucato2013}, extended in her work on mission-oriented policy~\cite{Mazzucato2018} and value theory~\cite{Mazzucato2018b}. Mazzucato argues that the state is not merely a passive corrector of market failures but an active co-creator of markets and technologies, and that publicly funded research constitutes a public investment whose returns should accrue to the public. Open knowledge infrastructure fits naturally into this framework: if public funding produces the research, the state should invest in the infrastructure required to make that research findable, usable and assessable as a public good. This argument has significant force. State investment has produced genuine public goods---from identifier infrastructure (ORCID, ROR) to national open science initiatives---and the Mazzucato framework provides the strongest theoretical justification for continued public investment in the open core of the annulus.

Moreover, the question of how to track and measure the returns on public investment in research infrastructure---which Lane, Owen-Smith and Weinberg~\cite{LaneOwenSmith2024} have recently addressed in the context of AI---is itself dependent on the kind of structured, linked metadata that the annulus model describes. The Entrepreneurial State cannot assess whether its investments are generating public value without the data infrastructure to measure outcomes, creating a recursive dependency: the state needs structured data to justify its investment in structured data.

But applying the Entrepreneurial State Model (ESM) to the \emph{full spectrum} of data refinement needs creates difficulties that the annulus framework makes visible. The welfare framework in Sec.~\ref{sec:optimal-width} makes the trade-off explicit: the ESM corresponds to the limiting case in which public subsidy $S$ fully replaces annulus revenue and the annulus width $w$ approaches zero. This is logically coherent but practically demanding for three reasons.

First, \emph{cost allocation}. In an idealised scenario, public institutions could centralise the refining process so as to avoid inefficiency and duplication of effort. However, in bearing the full cost of refining data to the quality levels required by pharmaceutical companies, venture capital firms and corporate R\&D departments, the public subsidy flows disproportionately to private beneficiaries. This is not market creation---the canonical justification for the Entrepreneurial State---but public subsidy of private competitive intelligence~\cite{Mingardi2015}. The distinction matters: Mazzucato's argument is strongest when the state creates infrastructure that enables private innovation (roads, internet protocols, basic research); it is weaker when the state produces the specific refined products that private actors would otherwise pay for.

Second, \emph{fiscal fragility}. State funding is subject to political cycles and fiscal pressures, creating structural vulnerability. A knowledge infrastructure entirely dependent on public funding is exposed to exactly the kinds of budgetary shocks that the Mazzucato framework seeks to prevent in other domains. The history of state-funded data infrastructure is not reassuring: databases have been defunded, privatised, or allowed to decay when political attention shifts. The annulus model suggests a more resilient arrangement in which the open core is sustained by a combination of public investment and revenue from commercial frontier activity, diversifying the funding base rather than concentrating it in a single source.

Third, \emph{geopolitical risk}. In a multipolar geopolitical environment, state-funded knowledge infrastructure carries risks of what Jasanoff~\cite{Jasanoff2005} has termed divergent ``civic epistemologies''---different political cultures' ways of establishing what counts as reliable knowledge. A knowledge commons anchored to any particular state, or even a coalition of like-minded states, risks encoding particular epistemological assumptions into what presents itself as a universal infrastructure~\cite{Czepan2024}. The concentration of infrastructure investment in the Global North has already produced a scholarly record with significant geographic biases; extending state-funded provision without addressing these structural biases may intensify rather than resolve the problem.

The annulus framework suggests a middle path: the Entrepreneurial State model is the right approach for the open core (foundational metadata, identifier infrastructure, the baseline that all institutions need), while the annulus provides the economic space for frontier refinement that serves differentiated demand and whose costs should not be socialised across the public purse. The governance challenge is ensuring that the boundary between these two zones is set in the public interest rather than by market power alone.

\subsection{The Crossref mechanism}

An alternative governance mechanism---one that has received less theoretical attention than it deserves---is already operating in part of the scholarly data ecosystem. Crossref, orignally as an industry collaboration among publishers and now expanding to be more inclusive in its membership, functions as a de facto boundary-setting institution for one zone of the annulus. When Crossref expands the scope of standard metadata disclosure---requiring, for example, that deposited records include reference lists, abstracts or ORCID identifiers---it effectively moves a data type from the annulus into the open baseline. This does not happen through state mandate or market competition but through collective agreement among the publishers of the underlying data.

This is a genuinely distinctive institutional mechanism. It is not state provision, not market dynamics, and not community self-organisation in the Ostrom~\cite{Ostrom1990} sense. It is collective disclosure by an organised industry body, and it has been remarkably effective at expanding the open core for data types that publishers already produce. Crossref has an increasingly diverse membership including publishers, research institutions, funders and governmental organisations. But the observation that Crossref has, without anyone having designed it for this purpose, become a boundary-setting mechanism for one part of the annulus raises an important question: under what conditions might similar mechanisms emerge for other data types? And might Crossref itself, with an appropriately expanded governance structure, play a broader role?

\subsection{The Barcelona Declaration as a norm-setting forum}

The Barcelona Declaration~\cite{BarcelonaDeclaration2024,KramerNeyWalt2024} introduces a governance logic that is compatible with the functional annulus model. Rather than mandating specific data to be free, the Declaration frames openness as a civic obligation---what Porter~\cite{Porter2024,Porter2026} has called ``research information citizenship''---and distributes responsibility across producers, consumers and aggregators of metadata. Waltman~\cite{Waltman2020} has argued that responsible research assessment \emph{requires} open scholarly metadata---a position that the annulus framework refines: the question is not whether metadata should be open in principle, but which metadata types should be inside the open core at any given stage of the system's maturation.

This framing is significant because it positions the Declaration not as a regulation but as a \emph{norm-setting body}. In the same way that the scholarly community has developed informal norms about PhD embargoes and data sharing, the Barcelona Declaration could be the venue where the community develops norms about which data types should be inside the open core, what quality standards apply to open metadata, and what responsibilities different actors have in the production chain. The Declaration's membership includes funders, institutions, and infrastructure providers---arguably the right constituency to develop these norms, though one could ask whether it should expand to include the corporate research users whose data needs shape part of the annulus.

Assessed against Ostrom's~\cite{Ostrom1990} design principles for commons governance, the Barcelona Declaration framework has both strengths and gaps. It establishes boundaries and articulates proportional responsibilities. But it lacks effective monitoring of compliance, graduated sanctions for non-compliance, and formal conflict resolution mechanisms. The history of analogous declarations---the Budapest Open Access Initiative, DORA, Plan~S---suggests that declarations without enforcement often fail to change institutional behaviour at scale~\cite{Hicks2015,Wouters2019}. However, it is not necessary for there to be explicit, direct enforcement associated with these initiatives for them to be of value.  Rather it is the engagement created by these approaches that engender change and new consensus in the policy environment. These initiatives do have each led to policy changes at institutional, local and national levels.  Thus, ensuring that they provide inclusive mechanisms to host debate and strength their arguments to make them more potent in the policy arena would seem to be an important facet of their function.

\subsection{Open access as a parallel case}

The open access experience provides a brief but instructive parallel. Open access mandates attempted to collapse the annulus in scholarly publishing by requiring that publicly funded research be freely accessible. What followed was not the elimination of commercial logic but its displacement: article processing charges preserved publisher revenues while shifting their form, and in some analyses total costs to the research community increased, with the burden redistributed regressively~\cite{pinfield2016,Khoo2019,Debat2020,Butler2023}. The annulus did not disappear; it migrated from access charges to publication charges while its structural function remained intact.

The annulus framework predicts this outcome: as long as there are genuine costs in the publication production system that exceed what can be funded through public subsidy alone, an annulus of some form will persist. The lesson for research metadata governance is that policy which treats the annulus as a pathology produces displacement rather than resolution. The functional annulus model avoids this trap by accepting the structural reality and focusing governance energy on the boundary conditions rather than on the elimination of commercial activity within the annulus.

\section{Dimensions: An Illustration of Annulus Dynamics}
\label{sec:dimensions-case}

Dimensions, the bibliographic database operated by Digital Science, provides an empirical illustration of annulus dynamics in practice. We describe it here not as a model to be copied but as a case study that makes several of the paper's theoretical claims concrete.

Dimensions was explicitly constructed on the foundation of open scholarly infrastructure---using Crossref DOIs, ORCID researcher identifiers and open metadata as its backbone, augmented by AI-driven entity resolution, classification and data enrichment~\cite{HookPorterHerzog2018}. The central thesis of its design was that by linking publications, grants, clinical trials, patents and policy documents into a single interlinked graph, it could provide a broader context for research than the traditional publication-citation ecosystem alone~\cite{HookPorterHerzog2018,HookPorterDraux2021}.

From its inception, Dimensions pursued a strategy of progressive data democratisation. A free version was made available in 2018 on the principle that researchers should be able to search the scholarly record without charge, and that analyses used in research evaluation should be reproducible against accessible data~\cite{Herzog2020}. Subsequently, the full dataset was made available on Google BigQuery, democratising not only access to data but access to the computational capacity required to analyse it at scale~\cite{HookPorter2021}. The rationale was that the combination of accessible data and on-demand computation could lower barriers for researchers, analysts and policymakers who had previously been excluded from large-scale bibliometric analysis by the cost of both data and infrastructure.

Each of these steps moved the inner boundary of the annulus outward for the academic research community. But Dimensions also maintains commercial products built on higher-order data refinement---institutional analytics, research landscape mapping, funder intelligence, patent-publication linkage---that serve the differentiated demand described in Sec.~\ref{sec:differentiated-demand}. The same underlying scholarly record serves radically different user communities at different quality levels. The commercial revenue from frontier refinement supports the infrastructure costs of maintaining the open and freely available layers.

An earlier episode in Digital Science's history illustrates the annulus lifecycle for a single data type with particular clarity. In 2015, Digital Science created the Global Research Identifier Database (GRID)---a comprehensive, curated database of research organisations worldwide, designed to solve the institutional disambiguation problem that the paper identifies as a key production friction. GRID was built because no open, community-governed organisational identifier system existed at the time, and Dimensions needed reliable institutional resolution as part of its data backbone. In December 2016, Digital Science released GRID under a Creative Commons CC0 licence---placing the entire dataset in the public domain without restriction. GRID was subsequently made available as Linked Open Data~\cite{Szomszor2020} and grew to cover over 100,000 institutions. Then in 2021, once the Research Organization Registry (ROR) had built sufficient community support, coverage, and governance maturity, Digital Science discontinued public releases of GRID in favour of ROR---effectively passing the torch to a community-governed identifier system that GRID had helped to catalyse. This trajectory traces the complete annulus lifecycle for organisational identifiers within a single organisation's experience: Frontier investment (creating GRID), deliberate intervention to push the inner boundary outward (CC0 release), and eventual transition to community governance (ROR) when the community was ready to sustain it. It also illustrates the structural constraint that open provision depends on prior frontier investment: ROR exists in its current form because someone first bore the cost of building the comprehensive organisational dataset from which the community effort could grow.

This practical experience illustrates several of this paper's theoretical claims. First, that the annulus has sectoral dimensions: different data types and different quality levels occupy different positions, and the inner boundary moves outward at different rates for different user communities. In terms of Sec.~\ref{sec:geometry}, Dimensions' trajectory represents a case where the inner arc has been deliberately pushed further out than market forces alone would produce---a strategic choice to increase the openness ratio for academic metadata in the annulus. Second, that the annulus is shaped by technological capability: as AI-enabled structuring lowers the cost of basic normalisation, data types that were previously commercially valuable become candidates for open provision. Third, that the boundary between open and commercial is not static but is determined dynamically by the interaction of production cost and user demand---and that a commercial actor can, by strategic choice, accelerate the outward movement of the inner boundary. Fourth, that the constraint identified in Sec.~\ref{sec:geometry}---that the inner boundary cannot move outward faster than the outer boundary---operates in practice: the open layers of Dimensions are sustained, in part, by the investment in frontier refinement that drives the outer boundary forward.

The limitations of this model should also be noted. The governance of the boundary between open and commercial layers is determined by the commercial actor rather than by community oversight. Open provision is contingent on the commercial operation remaining viable. And the reliance on a single commercial actor for a significant portion of open data infrastructure creates concentration risks. These limitations reinforce this paper's argument that governance norms of the kind the Barcelona Declaration is beginning to develop are needed to ensure that the boundary is managed in the public interest and not according to commercial logic.

\section{Open Questions}
\label{sec:research-agenda}

The framework developed in this paper raises several questions that define a research agenda at the intersection of science policy, innovation economics and knowledge governance.

\textbf{Optimal annulus width.} The welfare framework in Sec.~\ref{sec:optimal-width} yields qualitative predictions about how the optimal annulus width should vary by data type, but making these predictions quantitative requires empirical estimates of the functional forms: the social benefit function $B(r_i)$, the production cost function $C(r_o)$, the maintenance cost $M(r_i)$, and the revenue function $R(w)$. None of these is currently known for any data type in the scholarly ecosystem, and their estimation presents significant methodological challenges---not least because the ``radial'' dimension of the annulus is not directly measured in any existing dataset. Developing proxy measures---for example, using the time lag between commercial and open availability of specific metadata types as a proxy for annulus width, or using infrastructure provider cost data to estimate $M$---would ground the framework quantitatively. The dynamic extension, casting the problem as an optimal control problem with technological change driving the evolution of both boundaries, would connect to the literature on optimal technology diffusion and would yield predictions about the trajectory of annulus compression that could be tested using longitudinal data.

\textbf{Efficiency measurement and the openness ratio.} Can the distance between the actual scholarly data production system and the efficient benchmark be measured? The openness ratio introduced in Sec.~\ref{sec:geometry}---the ratio of inner to outer radius for each data-type segment---provides one candidate metric. Tracking this ratio over time for specific data types (citation data, institutional affiliation, CRediT metadata, funding acknowledgements) would yield an empirical measure of how quickly the open core is expanding relative to frontier structuring. Additional proxies include the proportion of metadata fields that require downstream normalisation, the rate at which standards adoption reduces processing costs, and the volume of duplicated AI-derived enhancement across the system. Decomposing annulus width into its technical and legal components, as discussed in Sec.~\ref{sec:emh}, would further refine the analysis.

\textbf{Governance mechanism for the inner boundary.} What institutional mechanisms could ensure the inner boundary moves outward appropriately? The Crossref mechanism works for data types already in the publishing workflow. The Barcelona Declaration provides a normative framework, but Ostrom's~\cite{Ostrom1990} design principles predict that governance without effective monitoring and graduated sanctions will erode under pressure. Empirical study of signatory behaviour change---analogous to the compliance studies conducted for DORA~\cite{Wouters2019}---would test the Declaration's effectiveness.

\textbf{Metadata provenance standards.} The distinction between principled and performative openness requires a formal metadata provenance standard. Existing foundations---the W3C PROV standard, DataCite's provenance model, Crossref's governance framework---could serve as building blocks. A standard analogous to verified carbon offsets would provide a more transparent basis for assessing the reliability of structured metadata.

\textbf{Equity implications.} How does the current annulus structure affect institutions in under-resourced settings? What level of open baseline provision is required to ensure that all institutions can participate meaningfully in the research ecosystem? These are empirical questions that could be addressed through comparative studies of data access and analytical capability across institutions of different resource levels.

\textbf{Cross-domain generalisability.} Similar annulus dynamics appear in clinical data, environmental monitoring, geospatial data commons and AI training dataset governance. Comparative analysis would test the model's generality and refine its parameters.

\section{Conclusion}
\label{sec:conclusion}

The persistent framing of scholarly knowledge infrastructure as a contest between openness and commerce has obscured a more productive question: how should the boundary between open and commercially refined data be governed so that the system as a whole serves the public interest?

This paper has argued that the innovation annulus---the zone between the open core and the advancing frontier of refined knowledge products---is not a pathology but a functional and permanent feature of the knowledge ecosystem. It exists because the cost of producing and refining structured knowledge data is real and persistent, shaped by production frictions that technology reduces but cannot eliminate. It is sustained by differentiated demand from communities---competitive institutions, research translation partnerships, corporate R\&D---whose quality requirements exceed what open provision can deliver. And its width for any given data type is a measure of the system's distance from perfect production efficiency: a distance that shrinks as standards mature and technology advances, but that is continuously regenerated as new data types emerge at the frontier.

Although we have principally confined our attention to consideration of scholarly metadata, we believe that this model can also be useful in thinking about the evolution of open access, and the commercial software / open source ecosystem where, with the rise of vibe coding, the following comments on AI are particularly pertinent.

AI has changed the parameters of this system profoundly. It has lowered the cost of basic structuring, pushing the inner boundary outward and expanding the volume of data that can be provided openly. But it has also raised the quality threshold at which refinement has value, shifted the commercially relevant frontier to higher-order tasks, while simultaneously introducing systemic risks through unprovenanced AI-derived metadata. Again, the annulus persists---not through resistance to openness, but because the economics of data refinement in a system with technological frictions and differentiated demand make it a structural feature of the landscape.

The governance question is how to ensure the annulus is the right width. The efficient market analogy suggests a benchmark: the annulus should reflect genuine production costs and nothing more. The welfare framework developed in Sec.~\ref{sec:optimal-width} makes this more precise: the optimal width for any given data type is determined by the interaction of the social benefit of openness, the cost of frontier production, the cost of maintaining the open core, and the financial sustainability of the system as a whole. Where the annulus has historically been inflated by market power and institutional lock-in---as in the Web of Science / Scopus duopoly era---competition and standards adoption have compressed it toward a more efficient width. Where it persists at the frontier---in domain-specific enrichment, integrity signals, and emerging metadata types---it reflects genuine investment that the system needs someone to make. Critically, the framework shows that an annulus compressed below the sustainable level does not merely reduce commercial revenue---it slows the advance of the outer boundary and, with it, the growth of the open core.

The Barcelona Declaration, with its framework of research information citizenship and its constituency of funders, institutions and infrastructure providers, represents the most promising forum for developing the community norms that should govern the inner boundary. It could, with appropriately expanded governance, become the venue where the scholarly community develops a shared understanding of which data types should be inside the open core, what quality and provenance standards apply, and what responsibilities producers, consumers and aggregators of metadata owe to each other. Whether it will develop the institutional machinery to make this vision durable---monitoring, enforcement, conflict resolution---remains to be seen.

The annulus will persist---the width is the price of being able to trust the data to the level needed to meet a given use case. The question is whether we govern it wisely: ensuring that the inner boundary moves outward as technology matures, that the open core is built on principled rather than performative foundations, and that the scholarly data ecosystem serves all its users---including those who cannot afford to pay for frontier refinement---as equitably and efficiently as the state of the art allows.

\begin{acknowledgments}
The author wishes to thank Mark Hahnel for valuable suggestions around Sec.~\ref{sec:optimal-width}, and Bianca Kramer, Simon Porter, Ludo Waltman, and Juergen Wastl their careful reading of this manuscript.  Any remaining errors or omissions lay solely with the author.

\textbf{Conflict of Interest Statement.} The author is CEO of Digital Science, a technology company that operates within the scholarly knowledge infrastructure landscape analysed in this paper. Digital Science's portfolio includes Dimensions, a bibliographic database that occupies a position within the annulus as defined herein, as well as Altmetric and other research analytics products. The author's position within this landscape informed the analysis but also constitutes a potential conflict of interest: the framework developed here could be read as a justification for the business model of the author's employer. We have sought to present the structural dynamics as objectively as possible, and we note that practitioner knowledge of the actual economics of data production is precisely what has been missing from much of the largely academic literature on open infrastructure. Readers should nonetheless be aware of this positionality when evaluating the arguments presented.
\end{acknowledgments}

\bibliography{market_dynamics}

\end{document}